\documentclass[runningheads]{llncs}
\usepackage{graphicx}
\usepackage{times}
\usepackage{listings}
\usepackage{dafny}

\begin{document}
\title{More Programming Than Programming: Teaching Formal Methods in a Software Engineering Programme}
\titlerunning{More Programming Than Programming}
\author{James Noble\inst{1}\orcidID{0000-0001-9036-5692} \and
David Streader\inst{1} \and
Isaac Oscar Gariano\inst{1}\orcidID{0000-0002-4881-0999} \and
Miniruwani Samarakoon\inst{1}}
\authorrunning{J.\ Noble et al.}
\institute{School of Engineering \& Computer Science\\
  Victoria University of Wellington\\
\email{kjx@comp.vuw.ac.nz}\\
\url{ecs.vuw.ac.nz/~kjx}}
\maketitle              % typeset the header of the contribution
\begin{abstract}
Formal methods for software correctness are critical to the future of
software engineering --- and so must be an essential part of software
engineering education. Unfortunately, formal methods are often
resisted by students due to perceived difficulty, mathematicity, and
practical irrelevance. We redeveloped our software correctness course
by taking a programming intensive approach, using the solver-aided
language Dafny to provide instant formative feedback via automated
assessment. Our redeveloped  course increased
student retention and resulted in the best evaluation for the course
for at least ten years.

\keywords{Formal Methods  \and Software Engineering \and Education
  \and Dafny.}
\end{abstract}

\section{Introduction}

In the last 20 years, formal methods for software verification have
moved from an esoteric research topic~\cite{rustan1998extended} to a set of
increasingly practical tools, and from doctoral study
to undergraduate degrees.  Victoria University of
Wellington’s Computer Science and Software Engineering programmes
include a course, SWEN324 “Software Correctness” that teaches software
verification.
We often call this course “Programming Made Hard” because 100 students
repeat the assignments they completed years ago in introductory
programming courses, but now must specify those programs’ behaviour 
and verify that their implementations meet those specifications.
In 2020 we redesigned SWEN324 using the solver-aided Dafny language,
supported by Leino’s Dafny textbook~\cite{dafny2020}; we are just
finishing teaching the 2021 version of the course at time of writing.
Students and teaching staff found the use of Dafny very positive: the
2020 course offering received the highest overall evaluation for at
least ten years.

Although very positive overall, students found Dafny 
difficult to learn and to use, and our informal observations as
teachers are that many of these difficulties stem from “accidental”
complexity introduced by the Dafny tool. This accidental complexity
obscures the “essential” complexity of learning the fundamentals of
software verification, and then applying those techniques to verifying
simple programs~\cite{Brooks1987}.
In this paper we reflect on our experience teaching SWEN324, focusing
particularly on our course design and issues with formal tooling.
%% The rest of this paper is as follows:\ldots

\section{Background}
% \paragraph{Tools for Program Proof}
Formal verification of software systems has been a significant
research topic for in computer science for 50 years or more~\cite{Jones2021}.  Tools such as Dafny, SAW, SPIN are increasingly
mature enough to support industrial application~\cite{Greengard2021,wayne2018temporal} but the main barrier to adoption remains a lack of software
engineers trained in their use~\cite{Garavel2020}.
To address this problem, there have been a number of studies on the
usability of formal methods, and tools that support formal
verification.  Beckert and
Grebing~\cite{Beckert2012} for example used the Cognitive Dimensions framework~\cite{Green1996} to
evaluate the usability of the KeY proof tool; Grebing and Ulbrich~\cite{Grebing2020} followed this up with a user study.

Tools have also been (re)designed to better support programmers in the
task of verifying their programs.  Whereas the Dafny tool, although
interactive, requires programmers to verify their whole program
statically, Gradual Dafny~\cite{Figueroa2018} allows programmers to choose
between static (“assert”) and dynamic (run time “assume”) verification
for each invariant.  Other gradual verification approaches have shown
similar promise at partial verification, but with choices embodied in
the tools themselves~\cite{Arlt2014,Pearce2015,Bader2018,Wise2020}.
Coming at the problem from the other side, Müller \& Ruskiewicz~\cite{Muller2011}
demonstrated how standard program debuggers could be used to debug
verification failures, by generating a modified program that
reproduced the failure when run, and Christakis~\cite{Christakis2016} integrated
concolic testing tools and lower level solver debuggers into Dafny’s
IDE.

More recently, some of the most interesting recent program verification work
has been using the Rust language~\cite{bornholt2021using}. Eschewing garbage
collection, Rust has an ownership types system that is used to manage
memory allocation, object lifetimes, and permissible inter-object
references.  Program verification tools such as Prusti~\cite{Astrauskas2019} and RustBelt~\cite{Jung2017} leverage ownership
information to support verification without needing memory structures
to be described separately.

% \paragraph{Teaching Formal Methods}

Finally, as formal methods' industrial use has increased, so has their relevance to education~\cite{Dongol2019,ferreiraformal,kamburjan2021increasing,FMFUN19};  Zhumagambetov~\cite{ZZZ} offers a relatively recent systematic literature review. 
Aceto and Ingolfsdottir~\cite{Aceto2021}, for example, have described a recent course at 
 the University of Reykjavik, where students can participate in a three week intensive formal methods course at first year. Yatapanage~\cite{Yatapanage2021} describes a recent second year course taught at De Montfort University that applied formal methods to concurrent programming --- although the paper's title highlights most students' concerns when approaching this topic ``\textit{Students Who Hate Maths and Struggle with Programming}''.  Kamburjan and Gratz~\cite{kamburjan2021increasing} showed how a custom interactive proof tool can generate a positive effect on student engagement; K{\"o}rner and Krings\cite{reliance2021} describe how pedagogical changes to inquiry-based learning can support the user of formal tools. 
In some ways closest to the approach we present here,
Ettinger describes how Dafny has been used for six years at Ben-Gurion University to support teaching refinement-style ``correct-by-construction'' programming \cite{teachDafny2021}, and 
Blazy describes a similar course based on Why3 \cite{why3teaching2019}.
G{\"u}demann describes how verification tools can even support similar learning strategies even in applied computer science courses taught using C \cite{teachFRAMA2021}.

\section{SWEN324 Software Correctness}

Formal methods have been taught as part of Computer Science and
Software Engineering programmers at VUW since 1984.  Unfortunately,
formal methods are often resisted by students due to perceived
difficulty, mathematicity, and practical irrelevance --- and SWEN324
had similar problems.
In 2021 we had the
opportunity to redevelop the course, as a companion to a relatively
new course SWEN326 ``Safety Critical Systems'', that focused on
correct software engineering in a wider context, including software
processes, testing, and abstract modelling (based on Alloy).  This
meant that we were able to refocus SWEN324 specifically on formal
methods for software correctness based on program proof.

\paragraph{Traditional Formal Course:}
We initially considered staying with a relatively straightforward,
``traditional'' formal methods course, introducing students to
propositional and predicate logic, then working up through weakest
preconditions to Hoare logics and their application in describing and
reasoning about software systems, culminating in pencil-and-paper proofs.  After some debate, it was decided
that this was not appropriate for several reasons.  In particular, our
students have already taken compulsory courses including Boolean
algebra and logic (as mathematics) and discrete logic (as physics)
during first year: we do not want students to regard this as ``another
maths or physics course'' --- our earlier experience with such courses
suggested that such a course would not be popular~\cite{formed08}. 
%%was left out, turned back on KJX
On
the other hand, our programme is heavily based around programming,
with all engineering majors requiring a full first year computer
science programme, and software engineering majors keen to take
practical elective courses to develop programming skill and
experience~\cite{formed08,Pang2018}.

\paragraph{Abstract Formal Modelling:}
We also considered taking an approach based on abstract formal
modelling.  High-level tools, such as TLA+~\cite{tlabook}, Alloy
Alloy~\cite{alloybook} or SPIN~\cite{spinbook}, support reasoning and
mechanised checking of systems' properties, based on abstract models
of those systems, rather than actual programming and source code.  It
is clear that these kinds of abstract formal models can play an
important role in software engineering projects, at least in
project's the early stages, supporting design validation before a
single line of code has been written.
Indeed,
we had earlier taught a first-year course (SWEN102) that attempted to
give a gentle introduction to formal methods within a more general
context of software modelling, beginning with UML and moving through
to Alloy~\cite{formed08}.  The idea was to present formal and informal
approaches as different points in a spectrum of approaches to
describing software systems, rather than being totally different
subjects.  We also wanted to ensure that students see software
modelling as a useful way of understanding systems, rather than just
an exercise in learning new notations, so we felt it was important
that any formal notation we used be supported by tools which allowed
students to explore the consequences of the models they created.  This
course was mostly successful on its own terms: even first year
students were generally capable of domain modelling using Alloy,
of translating functional requirements into Alloy properties, 
and then able to analyse the Alloy models to demonstrate that the
requisite properties held (or explain why they did not).

Unfortunately, our SWEN102 course was never widely popular: for better or
for worse, our cohort, privilege programming, over pretty much every
other software engineering activity or practice. For the SWEN102
approach to work, we first had to successfully ``sell'' modelling, and
then second to ``sell'' the advantage of formal models over informal
ones --- where students simply did not see the relevance of the models
to the programming / software engineering tasks the expected to
undertake.  On the other hand: SWEN102 demonstrated that even our
early undergraduate students were capable of learning formal tools,
constructing formal models, and handling propositional and predicate logic.

\paragraph{Formalism as Programming:} 
For this reason, we decided to base our SWEN324 course redesign on the
reverse of the traditional approach.  Rather than progressing bottom
up from propositional logic to predicate logic, Hoare logic, and
eventually perhaps experimenting with a practical tool, we aimed to
progress top down: starting with programming language based tool, and
then using that high-level tool as a context in which we can present
and teach the key concepts of software correctness --- while offering
the majority of students an experience that feels like programming,
rather than like doing mathematics.  

The latest version of this course --- SWEN324 ``Software Correctness''
-- adopted the Dafny programming language and associated toolset,
based on the Z3 solver and the Visual Studio Code.  
Dafny provides what Leino has called ``auto-active'' verification \cite{autoactive2010} 
in which verification is seamlessly incorporated into development practices
and the toolchain. It may be clearer to think of this approach as \emph{implicit} verification 
where programmers annotate their programs with preconditions,
postconditions, variants, invariants, as in Eiffel \cite{touchOfClass},
and do not interact directly with formal models or e.g.\ proof trees. 
This is in contrast to \emph{explicit} verification technologies
such as Coq \cite{CoqTute11,Chipala2013} where programmers must 
interact with solvers by directly building proofs and proof trees,
potentially even extracting programs from those proofs. 
Dafny's implicit approach still offers many guarantees:
Dafny attempts to prove programs totally correct by default,
so recursive methods and loops often require programmers to 
give variants to prove termination, and loops in particular 
generally require invariants to prove correctness. 
Array and pointer accesses typically require invariants, assertions,
or preconditions to ensure all accesses are within bounds and 
variables are initialised and non-null. This means that
Dafny programmers (and thus students)
interact with Dafny's underlying prover indirectly, at arm's length,  
in terms of definitions in their programs and constructs in the Dafny language,
rather than having to learn explicit representations of proof. 

\paragraph{Choice of Dafny:}
Dafny was selected
for a number of pragmatic reasons: it is well supported by a team in
Amazon’s Automated Reasoning Group led by Rustan Leino,  has
substantial publicly available on-boarding and tutorial material,
including a full book by Leino~\cite{dafny2020}, an online playground at
Rise4Fun, documentation available online, and a developing academic
community --- and, frankly, because what little experience 
the course staff had with suitable tools seemed most transferable
to Dafny.
Based on our earlier experience, we hoped Dafny would offer a number
of advantages over Alloy, or more sophisticated tools like Coq \cite{CoqTute11} or Why3 \cite{why3teaching2019}.  First, Dafny offers a concrete, ASCII-compliant
syntax --- being restricted to ASCII means students
should feel some familiarity with the notation:
students would not need to learn how to type, let alone
pronounce, relatively esoteric characters such as $\alpha$, $\delta$,
or $o$ (little were we to know how familiar alpha, delta, and
omicron would become). 
Dafny's syntax
and semantics being based on C$\sharp$ and Java should also be familiar. Students can use the development
toolsets they already know, such as VS Code, Eclipse, Git --- particularly important for students who need 
tools such as screen readers, magnifiers, or voice control
to complete their work.

Second, because Dafny is well supported by a toolset, we are able to
rely on Dafny itself to provide students rapid formative feedback ---
simply by requiring students to submit their solutions via the Dafny
verifier.  In a very real sense, we are able to leverage the ``essential
difficulty'' of formal verification of correctness --- that no only must students implement a correct program, but they must also convince the
Dafny prover that their implementation is correct --- to aid the
students in that task.  In simple cases, where students' focus on
implementing programs, we can directly supply students with the Dafny
specifications and the tool itself will provide feedback: either their
program verifies against the specification, or it does not.  Where
students' focus is on writing specifications, we can allow students to
verify their solutions against hidden ``oracle'' specifications, and
again Dafny can check that the students' specifications capture
important properties described by the oracles, or more
straightforwardly, that the students' specifications and the oracles
are mutually consistent~\cite{matrixOracle}.

Finally, because Dafny is relatively mature, there is a fair amount of
material available online, which students are able to access as
necessary. We were also able to use a draft version of Leino's
\textit{Program Proofs} textbook~\cite{dafny2020}. 

\paragraph{Continuous Automated Feedback:}
The ability for Dafny to provide feedback, and that this course was
targeted at third-year students --- experienced both in programming
and in tertiary study --- lead us to make this automated feedback a
central feature of the course.  Again based on our department's
practice in teaching programming --- with which our students are very
familiar! --- we provide that feedback in two ways.

First, our ``lectures'' are centred around a weekly series of small
``mastery'' questions about Dafny and verification, served from a
simple website.  This is similar to the
existing Dafny Rise4Fun website, but simpler: we discuss this further
in the next section.  The weekly questions are
released at the start of each week, and students may discuss the
questions, may work in groups, ask for answers, and make any number of
attempts at answering them --- but are expected to answer the vast
bulk of these questions correctly.  The time in ``lectures'' allows
students to discuss any of the questions with the class, lead by the
course staff --- in practice, the website lets us know which questions
students are currently finding difficult, and so we use that to guide
choices.  Because of the very liberal rules around answering the
mastery questions, we can work out the solution to any weekly question
in class, and even demonstrate the correct answer and show it
verifying: if students choose to pay little attention and just copy
the provided answer, so be it.

Second, we also incorporate automated feedback into larger summative
individual assignments (again, we provide examples in the next
section).  Students can submit answers to the assignments as many
times as necessary: by running each submission through the Dafny
verifier, students then get immediate feedback about their submission.
This feedback is quite terse (just the number of assertions verified,
or not verified) because it is not intended to replace students' use
of IDEs or to substitute for their own attempts at verification ---
rather it is so students can judge their progress through the course,
and in particular, to know when they have completed each part of each
assignment.  We are careful to ensure that every important concept
required by the summative assignments are covered by weekly questions
before the assignment is due.  Thus, while we can discuss the
summative assignments only in broad outline, we can (and do) refer
students to the relevant weekly questions which we can discuss in as
much detail and at as much length as necessary.

\paragraph{Course Design:}
As with all VUW engineering courses, SWEN324 is offered in one twelve week
semester, generally split into two six-week
half-semesters. Figure~\ref{courseplan} shows the ideal course plan
(for COVID reasons, an extra week's break was substituted at week 9 in
2020 and week 3 in 2021).   There are four main topics in the course:
learning Dafny as a programming language; writing Dafny (method) specifications;
verifying those specifications against Dafny programs; and handling objects with mutable state.

\begin{figure*}
\vspace*{-5mm}
\begin{center}\small
\begin{tabular}{|l|p{.53\textwidth}|l|}
\hline
Week & Topic & Assignment\\
\hline
\hline
1. & \textbf{Introduction} --- overview, industrial use~\cite{irondafny,Cook2018} &  \\
2. & \textbf{Programming} --- pre- and post-conditions &  \\
3. & \textbf{Data} --- inductive data types, pattern matching & A1 Programming (10\%) \\
\hline
4. & \textbf{Recursion} --- totality, termination   & \\
5. & \textbf{Structural Recursion} ---  over inductive types & \\
6. & \textbf{Iteration} --- loops, variants & \\
\hline
\hline
% &  & \\
& Midterm break & \\
& Midterm break & A2 Specifying Programs (15\%) \\
% & & \\
\hline
\hline
7. & \textbf{Loops} -- loop invariants and variants &   \\
8. & \textbf{Recursion vs Iteration} --- tail recursion & \\
9. & \textbf{Recursion vs Iteration} --- specs. vs.\ programs  & \\
\hline
10. & \textbf{Objects} --- mutable structures and validity & A3 Verifying Programs (15\%)\\
11. & \textbf{Ownership} --- ownership of representation &  \\
12. & \textbf{Proofs} --- Dafny ``\texttt{assert}'' vs ``\texttt{calc}'' & \\
\hline
15.  & & A4 Reasoning about Systems (40\%) \\
\hline
%\hline
\end{tabular}
%\vspace*{-5mm}
\caption{SWEN324 course plan.}
\label{courseplan}
\end{center}
\vspace*{-10mm}
\end{figure*}

\paragraph{Course Content:}
The resulting course covers most of the content Leino's \textit{Programs Proofs}~\cite{dafny2020},
although it does not explicitly address the foundational material.
In more detail: we address essentially all the ``core'' features of Dafny circa 2020, i.e.\ Dafny version 2.3.0. This included Dafny methods and classes (imperative, and mutable); functions and inductive datatypes (immutable, finitary); pre and postconditions; predicates (Boolean functions); assumptions and assertions; compiled vs ghost code, well-founded recursion and explicit termination measures, pattern matching, destructors; 
built-in collections (arrays, sets, maps); loops, invariants, and variants; 
recursive specifications of iterative programs (including transformations between general recursion, tail recursion, and iteration); and
representation invariants for dynamic data structures.

There are only two chapters of material from \emph{Program Proofs} that we intentionally overlook.  Chapter~2 presents the mathematical foundations of Dafny's program logic, based on Hoare Logic and Weakest Preconditions.  Where necessary, we discuss Dafny's semantics informally: we have not needed to refer the formal definitions.  Chapter~5 presents the notion of proof and Dafny's constructs (function lemmas, \texttt{calc} blocks) that can support programmers in making explicit proofs.  Perhaps more surprisingly we have not needed this material either. Because Dafny is an \emph{implicit} verification system, students do not need to build proof objects, and they are not even able to see what proofs Dafny's solver many have constructed!

\paragraph{Course Assessment:}
The overall assessment of the course is shown in Figure~\ref{markstable}.
A significant fraction of the assessment supports the formative mastery questions, with
the balance taken up by four summative assignments, one for each part,
and a reflective essay.  Each part of the course is addressed by around
25 weekly formative mastery questions.  Students who complete all the
mastery questions and the first assignment are well on the way to
obtaining a bare pass; students who are hoping for an ``excellent''
grade must complete most of the assignments correctly.

\begin{figure}
\vspace*{-5mm}
\begin{center}
\begin{tabular}{lr}
Weekly Overview Questions (Dafny, open) & ~~~~~20\%  \\
Assignment A1: Programming (Dafny, individual) ~~~ & 10\%  \\
Assignment A2: Specifying (Dafny, individual) & 15\%  \\
Assignment A3: Verifying (Dafny, individual) & 15\%  \\
Assignment A4: Reasoning (20\% Dafny, 20\% essay, individual) & 40\%  \\
\end{tabular}
\end{center}
\vspace*{-5mm}
\caption{SWEN304 Assessment Items.}
\label{markstable}
\vspace*{-5mm}
\end{figure}

These assessment weights also guide students time.  VUW courses of this size (15 points) are
rated at 150 hours over the whole trimester --- nominally 10 hours
per week over 15 weeks --- 12 lecture weeks and a three-week
assessment period at the end. Allowing approx.\ 25 hours (2 hours per
week) to attend lectures, and another 25 hours for background reading,
installing software, navigating Git, etc, that leaves 100 hours of assessed work. 
The assessment percentages offer a rough guide to
the amount of time students should aim to spend on each piece of work.

%KJX was previously deleted, now restored.
\paragraph{Course Objectives:}
The resulting course objectives are that, by the end of the course,
students should be able to:

\begin{enumerate}
\item Explain what it means for a system to be correct, what engineering techniques we can use to increase confidence in correctness, and why this is important.
\item Use formal structures such as sets, functions, relations and sequences to model software systems.
\item Use formal notations to specify desired properties of software systems, such as assertions, pre- and postconditions, variants, and invariants.
\item Use formal tools to check that systems correctly implement their desired properties.
\item Use formal reasoning to explain why a particular system is correct with respect to a specification.
\end{enumerate}

The first objective is primarily tested by the essay: the other
objectives by the assignments and mastery questions.

\section{Assessment}

To quote Tom Angelo~\cite{TomAngeloWherever}, ``\textit{most students
  are going to try to `study to the test.'} ''.   What is assessed is
what we can expect students to (try to) learn. This is why we have
restructured SWEN324 around questions and assignments with automated
feedback, rather e.g.\ than traditional lecture content. 
In this section we present examples of the assessment items we
designed for SWEN324, to demonstrate the kind of problems students are
able to solve during the course.

\subsection{Weekly Overview Questions}

As discussed above, 20\% of the assessment in SWEN324 is in the form
of formative weekly questions.  Students can choose to answer any
question at any time, and make repeated attempts to answer each
question. The point is formative, to support learning, rather
than summative evaluation --- although
the system records when each student successfully answers
each question.  Students can repeat completed questions (e.g.\ to
experiment with alternative solutions) --- the question stays listed
as completed.

\begin{figure}[hbt]
\vspace*{-5mm}
\begin{center}
  \includegraphics[width=\textwidth]{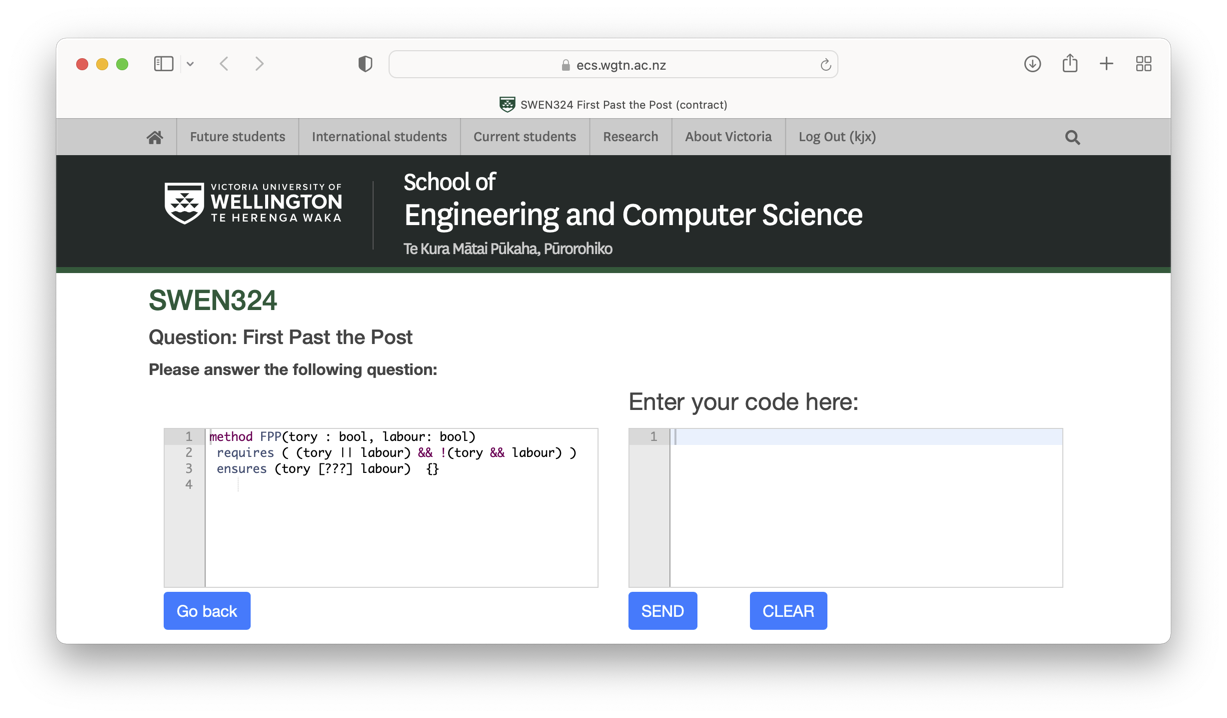}
\vspace*{-15mm}  
\end{center}
\caption{Web interface for weekly questions.}
\label{weekly}
\vspace*{-5mm}
\end{figure}

Figure~\ref{weekly} shows the
rudimentary web system that presents these questions to students.
The
left-hand pane shows some Dafny code including a place-holder
``\verb+[???]+''; this placeholder is replaced by whatever students
type in to the right-hand pane.  This system was originally built by
our colleague Marco Servetto to help students revise their Java
knowledge, and is well integrated with the other systems
which we use in the school: we have re-purposed this tool for
Dafny.

The question in Figure~\ref{weekly} (titled ``First Past the Post'')
is addressing a basic definition of Boolean algebra: what is Dafny's
Boolean ``exclusive-or'' operator. This question shows the advantage of
the placeholder mechanism: potential solutions are necessarily
restricted to fit within the syntactic context of the placeholder. The
solution to this question is Dafny's ``\verb+!=+'' operator.

\begin{figure}[hbt]
\vspace*{-5mm}
\begin{center}
  \includegraphics[width=\textwidth]{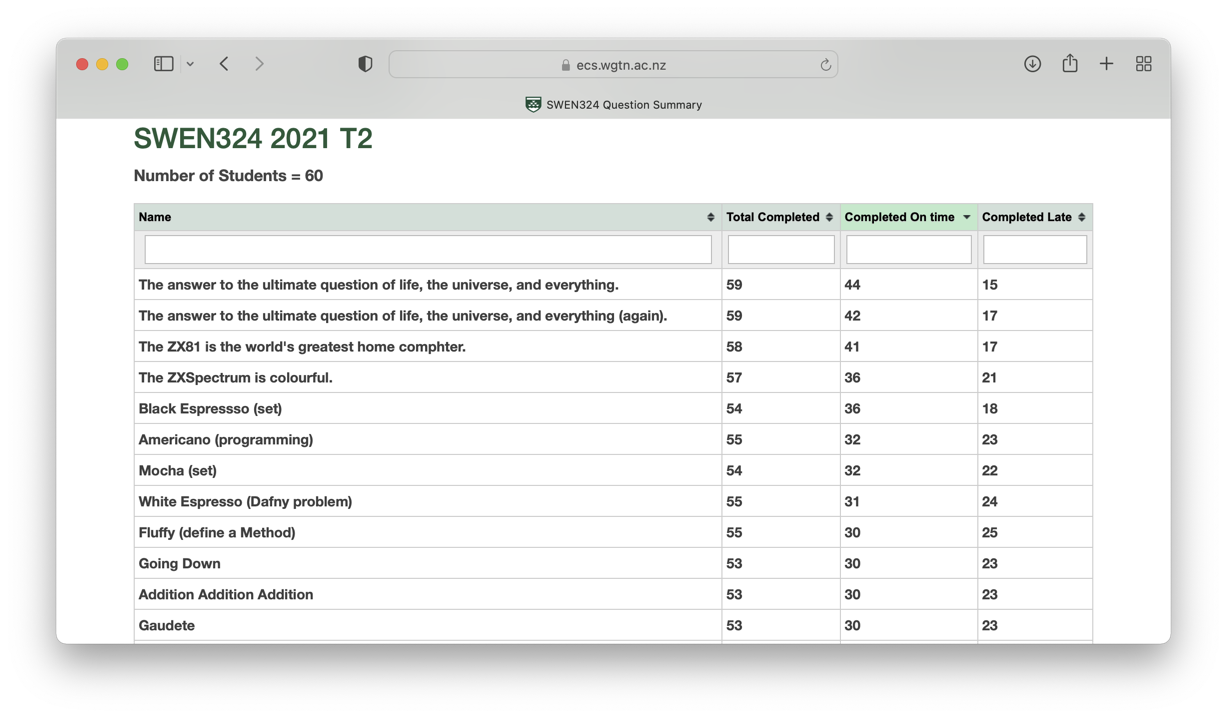}
\vspace*{-15mm}  
\end{center}
\caption{Overview of student progress.}
\label{overview}
\vspace*{-5mm}
\end{figure}

Figure~\ref{overview} shows the course-wide overview of the summary
questions, showing how many students have completed each
question. This proved very used in tracking students' progress
through the course overall, and in choosing lecture topics
(i.e.\ which questions we will discuss and then answer in lectures).
Generally we aim to pick questions where that top 10-20\% of students
have answered successfully (we can lure them into the discussion of
their solutions) but the bulk of the class has not (so that they are
interested in learning how to solve those questions).  This also
allows us to choose not to revisit questions that the vast majority of
the class has already answered, even if some stragglers have not ---
rather than taking up everyone's class time with well understood
topics.
Rather, we can direct stragglers e.g.\ to the recordings of the
lectures where we have answered those questions, or arrange to provide
individual support.

It is worth reiterating that these questions are at least as important
as resources or content or prompts for lecture sessions, as 
questions that students must answer by themselves.  Fairly early on,
for example, there is a relatively simple question 
that most students get wrong: 

\begin{lstlisting}
//complete the following method which returns the "real"
//sum and product of its two real arguments
method SumAndDifference(a : real, b : real) [???]

//Hint: https://www.youtube.com/watch?v=kqFPDrDWAHs
\end{lstlisting}

\noindent The point of this question is that the question title
(``We'll look at them together then we'll take them apart'')
and method name (``\texttt{SumAndDifference}'') are
inconsistent with the comment on the method
(``\texttt{//.. "real" sum and product}'').
This inconsistency was originally introduced in error,
however we kept it because of the valuable in-class discussion 
it engendered, about how comments can be misleading,
as can method names, or alternatively tests or specifications
can be incorrect. 
As it happens, here the comment is wrong: the automated
test indeed requires sum and difference not sum and product.

The ``First Past the Post'' question illustrates how we use
Dafny to revise Boolean algebra. The questions get rather
 more sophisticated as the course progresses. 
For example, the ``Very
Logical, Mr Spock'' question also tests Boolean algebra, but requires
students to understand how a method's control flow and assignments are
summarised by postconditions (``\lstinline+ensures+''):

\begin{lstlisting}
method logical(a : bool, b : bool, c : bool) returns (t : bool) 
  ensures [???]
{
 t := false;
 if (b) {
    if (a) { t := true; } }  else { t := false; }
 if (c) { t := a; }
}  
\end{lstlisting}

The ``How many leaves'' question requires students to write a
recursive function to calculate the size of a tree:

\begin{lstlisting}
datatype Tree = Leaf | Node(left: Tree, right: Tree)
function method Size(t: Tree): nat
[???]
        
method Main() {
  var tl:Tree := Leaf;
  var tc:Tree := Node(Node(Leaf, Leaf),Leaf);
  assert Size(tl) == 1;
  assert Size(tc) == 3;
  print "  ",Size(tl),"  ",Size(tc), "\n";

}
\end{lstlisting}

%\pagebreak  %%KJX

The ``Hopalong'' question requires students to define a termination
measure, as Dafny programs are total by default:
\begin{lstlisting}
//insert a decreases clause so Dafny can prove termination
function hopalong(q: int, x : int, y : int, z : int) : int
  [???]
{
 var modulo := (x + y + z) % 3;

q + if (y <= 0) || (z <= 0) || (x <= 0) then 0 else 
if (modulo == 0) then (hopalong(q+1, x + 3, y - 1, z + 2))
   else if (modulo == 1) then (hopalong(q+3, x - 3, y , z - 1))
   else (hopalong(q+5, x + 2, y, z - 10))
}
\end{lstlisting}

Our final example is an excerpt of the last of the weekly
questions --- the full example presents 90 lines of code to students;
another 30 lines of code for method implementations are omitted.  This
question is rather more complex, requiring students to implement both
the ``\lstinline+Valid()+'' predicate to describe the class invariant of a
complex mutable object, and to manipulate the ``\lstinline+Repr+'' ghost
field that must track the auxiliary implementation objects owned by
the stack:

\begin{lstlisting}
datatype StackModel = Empty | Push(val : int, prev : StackModel)

class Stack {
   var values : array<int>
   var capacity : nat
   var size : nat
   
   ghost const Repr : set<object>

//Define these two methods so that the hidden code below works
//  constructor(capacity_ : nat) 
//  predicate Valid()
[???]

   method push(i : int) 
     requires Valid()
     ensures Valid()
     modifies Repr
     ensures capacity == old(capacity)
/*omitted*/
\end{lstlisting}

%% \begin{tabular}{p{5cm}p{5cm}}
%%   %\begin{lstlisting}[language=dafny]
%%   %\begin{lstlisting}
%%     FOO1
%%     %\end{lstlisting}
%% %\end{lstlisting}
%% &
%% %\begin{lstlisting}
%% FOO2
%% %\end{lstlisting}
%% \\
%% \begin{minipage}{5cm}
%% \begin{lstlisting}
%% FOO3
%% \end{lstlisting}
%% \end{minipage}
%% &
%% %\begin{lstlisting}[language=dafny]
%% FOO4
%% %\end{lstlisting}
%% \\
%% \end{tabular}

\subsection{Assignments}

The four Dafny assignments are very similar to the overview questions
in spirit --- but with two main differences: they are undertaken using whichever Dafny IDE students choose (usually Visual Studio Code);
and students must upload complete Dafny files into the school's  standard
submission system, rather than using a specialised web interface. Assignment questions are significantly larger than weekly
questions.  Whereas the overview questions typically aim to teach
one single verification concept or Dafny construct, the assignments
typically require students to combine techniques and link concepts
together. To guide students' work, we again ensure
rapid feedback by reporting the results of Dafny attempting to verify
each submission, and we allow students to submit work any number of
times.
Space does not permit us to include full details of assignment
questions here --- however some of the more interesting questions included:

\begin{enumerate}
\item Add annotations to the code of a vector sum (A1) or small sorting
  network (A1).
\item Print out the text of the song ``Ten Green Bottles'' (exactly
 as supplied, 1743 characters) but with a program shorter than 750
 characters (A1).
\item Calculate the income tax payable by an individual New Zealander (A1).
\item Calculate with Carolingian duodecimal currency or interval
  arithmetic (A2).
\item Verify functional implementations of sets, lists, and maps (A2,A3).
\item Test if a string is a Palindrome (A4).
\item Implement search trees (A3), tries (A3), or balanced trees (A4).
\item Implement an object-oriented mutable map (A4).
\end{enumerate}

These questions obviously get harder as they go along. The first questions either ask students to annotate existing code, or write code without specifications to introduce students to the language. Even here, however, apparently simply programs such as ``Ten Green Bottles''
(which we do not verify against any external specifications) still require significant verification effort to be accepted by Dafny ---
at least four or five lines of annotation out of a 25-line solution.
Dafny needs to prove termination, and that all array accesses
are in bounds, and this necessitates preconditions constraining
arguments on all subsidiary methods and functions.   The final assignment questions are as complex as the final data structure examples from 
\textit{Program Proofs}.

\subsection{Essay}

%% \begin{verse}
%%   \textit{
%% "getting code to work is one thing.\\
%% proving it does what it's supposed to is something else.\\
%% convincing dafny you've proved it does what it's supposed to\\
%% is something else entirely.
%%     }
%% \end{verse}
%% \begin{flushright}
%% Thomas J. ``Tad'' Peckish (attrib.), twitter, Oct 4 2020 \\ ``motto for a software correctness course''
%%   \end{flushright}

A reflective essay provides the last 20\% of the course.  This is the
final assessment component that students complete --- although due to
VUW's regulations, it is due together with the fourth Dafny
assignment, as late as possible in the term.  The core rubric for the
essay is straightforward: to write no more than 750 words reflecting
on students' ``experience with verified programming in Dafny to ensure
software correctness', in the style of a blog post aimed to
communicate to other students, developers, or software engineers.
Students are invited to select a problem (typically from the final
assignment, but ``in case of emergency'' they may choose any
programming problem) and then explain how they used Dafny to specify,
implement, and verify their chosen problem; to discuss which features
of Dafny made this easier (or harder); and if they had to do it again,
what they would do differently and why.

This essay fulfills two important purposes in the course design.
Towards higher marks, a VUW ``A-'' grade is 80\%: a student who completes all the assignments perfectly
but chose not to attempt the essay would get that grade. The essay
thus enables us to distinguish the truly outstanding ``A+'' students from
the merely excellent ``A'' or ``A-: students.  At the other end of the grade
distribution, reasonable attempts at the weekly questions and the
first two assignments should yield 40\%:  an essay
that demonstrates merely ``adequate evidence of learning'' is then
sufficient to pass the course.

\section{Experience with Dafny}

Mathematics may still be taught via pencil and paper (or \LaTeX) but
these days teaching programming is impossible without a toolchain: a
language implementation, a development environment, and the other
accoutrements students expect.  Our course design teaches verification
as a specially intense kind of programming (``More programming than
programming is our motto'' \cite{bladeRunnerFilm}) --- this requires
a toolchain that is reliable, scalable, and supported enough to cope
with daily use by hundreds of students.
Luckily, we found the current versions Dafny were certainly good
enough for our purposes: we were able to spend the vast majority of
our efforts in teaching the practices and principles of verification,
rather than working around problems and bugs in the tools. While we
encountered roughly one serious bug during each course offering so
far, the Dafny project team resolved them assiduously.Our overall
experience with Dafny was very positive. 

Probably the biggest issue we
encountered was just finding the resources -- notably staff time and
effort --- to support rapid feedback via automated marking of the
weekly questions and the assignments. The problem was not so much the
necessary infrastructure, which is essentially a one-off cost, but the
advance preparation needed for automated marking of every assignment.
Basically, marking must be complete before an assignment can be
released, rendering it no longer possible to write underspecified assignments
which point students in a general direction, wait until the assignment deadline, and then take as much
time as necessary \textit{after the students have submitted their work}
to work out the marks, the desired solutions, or
even \textit{whether solutions are possible}. All this work must now be completed beforehand.

\vspace*{5mm}

That said, we did strike three more technical issues that could be
addressed via changes to Dafny's design:

\paragraph{Program Testing:}  We encourage students to start by testing
their implementations, because it is easier to verify code that is
correct
% against specifications that are also correct
than it is to
verify incorrect code :-).
Dafny’s tight integration of proving and programming unfortunately
means that programs cannot easily be tested until they are fully verified.
We observed students continually “commenting out” assertions and
preconditions to be able to test their programs, and then undoing
those comments to undertake verification. There are four related problems here.

First, Dafny's requirements to prove all memory accesses safe,
and that all programs terminate, often mean even simple programs
have to be heavily annotated just to compile.
A method to swap two array elements will require 
array reads and writes to be in bounds;
the obvious (and best practice) solution is
to define method preconditions which ensure method 
arguments are in bounds: but now all callers of the
method must themselves do enough to meet those preconditions. 

Second, while annotations, assumptions, and non-totality declarations etc.\ can be used 
to remove the need for some of these checks, they still require
students to annotate their programs explicitly, 
i.e.\ so students always have to deal with the checks
even if just to tell Dafny to ignore them!

Third, while Dafny does support command line options to e.g.\ 
ignore verification and compile and run programs directly,
verification is an all-or-nothing, static affair: either
verification is attempted for the whole program, or 
all specification and verification constructs are ignored.

Fourth (and finally) the options to control verification are buried in
the command line, and are not surfaced in the Visual Studio Code IDE.

Following the example of
Gradual Dafny \cite{Figueroa2018} and Gradual Verification 
\cite{Arlt2014,Bader2018,Wise2020}
more generally should make testing easier.
Ideally students would be able to run programs in a “test mode” where
Dafny checks as many assertions, assumptions, and pre- and postconditions
as possible dynamically.  Students could then express a series of unit tests as
Dafny assertions: if the program verifies, well and good; but if not,
they would still have the option of running the program and using
print statements or host debuggers to interrogate program state.
Recent Dafny releases \cite{Dafny3.0.0} now support an 
\texttt{expect} statement that does Gradual Dafny style dynamic
checking: implementing this option may be as simple as translating
Dafny's verification condition as \texttt{expect}s rather than \texttt{assert}s. 

\paragraph{Verification Debugging:}
Much of the work of verifying Dafny programs involves students
annotating their code --- adding require and ensure clauses and
assertions until the verifier has enough information to discharge its
proof obligations.  Students find this hard because it is not obvious
what Dafny “knows” at any given program point: which assertions Dafny
is able to prove, which assertions Dafny is able to refute, and which
assertions Dafny is unable to answer (i.e. where the prover times
out). We also observed cases where Dafny is unable to verify an
assertion because it does not have enough information about variable
values --— this is particularly prevalent in code where e.g.\ 
students have forgotten to write method postconditions, or 
have not realised a particular postcondition is necessary. This
manifests as Dafny being unable to verify an assertion about a
method’s return value, and simultaneously unable to verify the
negation of that same assertion.  Even good students find this
situation intensely frustrating.  Ideally Dafny would be able to give
programmers more information about what it knows, e.g.\
by querying its underlying solver \cite{Christakis2016}.

\paragraph{Mutable Object Structure:}  Dafny is one of the few tools
that can verify programs built from composite structures of mutable
objects using class invariants and representation sets.  In practice,
this requires either explicit definitions of ``\lstinline+Valid+'' and
``\lstinline+Repr+'' attributes \cite{dafny2020} which are verbose and
complex, or implicit definitions generated via the
``\lstinline+autocontracts+'' attribute \cite{leino2013} which are
concise but opaque.  Few students were able to use either mechanism
effectively.  Perhaps by building on work verifying Rust programs,
such as Prusti \cite{Astrauskas2019} and RustBelt \cite{Jung2017}, it
should be possible to add ownership annotations to fields and
parameters, to check those annotations as with Rust's borrow checker
\cite{Markstrum2010,Dietl2011,Klabnik2019} and thus extend the
implicit definitions already generated by autocontracts.

\vspace*{5mm}

We also encountered a number of pragmatic issues that arose with Dafny, but which appear to be consequences of Dafny's design choices, and as such are less amenable to technical fixes.

\paragraph{Idiosyncrasies:}  Dafny's syntax is sometimes idiosyncratic, which students found hard to follow.  
To give just one example, here are a method and function to add two numbers:
\begin{lstlisting}
method addM (a : int, b : int) returns (c : int) { c := a + b; }
function method addF (a : int, b : int) : int { a + b }
\end{lstlisting}
The syntax for declaring the return values are different
(\lstinline+returns+ vs \lstinline+:+); the syntax for actually returning the results are different; a final semicolon is mandatory in the method and forbidden in the function. 
Adding insult to injury, methods and functions then perform very differently in the verifier:
\begin{lstlisting}[numbers=left,stepnumber=1]
     var m := addM(x,y);
     var f := addF(x,y);
     assert m == x + y;   //Fails to verify
     assert f == x + y;   //Verifies
\end{lstlisting}
Dafny verifies the assertion on line 4, because functions are incorporated into the verification context.
Dafny fails to verify the assertion on line 3, however,
because methods are always abstracted by their postconditions,
and the declaration of \lstinline+addM+ omits postconditions.
There are reasons for these choices, but they do make the language more difficult to learn.  

\paragraph{Implicit vs.\ Explicit Verification:}  We have described as taking an implicit (aka ``auto-active''  \cite{autoactive2010}) approach to verification. 
Our students, or Dafny programmers in general, 
do not construct proofs explicitly, in some verification domain that reflects on the base domain of the program: rather they
work in an extended programming language domain. 
That is, students focus on programs, and program verification, but not on the foundations of logic, programming languages, and critically, not on proof.  Our teaching practice 
builds on this implicit approach: students 
definitely need an implicit understanding of the 
underlying formal concepts --- because they
will be incapable of completing 
any work without that understanding --- but we present 
those concepts completely within the programming 
approach: we don't discuss the semantics 
of programming languages, weakest preconditions, 
the kind of inferences Dafny's underlying solver is making,
let alone how it works.
We approach software verification in
the same way that most software engineering courses 
approach statically-typed languages: students can
understand the benefits, and use the type systems,
but could not give a type-theoretic explanation
for why their programs don't compile.

Arguably the biggest weakness of this implicit approach
is that it sidesteps the question of proof.
Dafny does not illustrate proofs of programs
(other than symbolic dumps designed for debugging Dafny).
As a result, we do not expose students to formal proofs,
and in fact students never need to understand what a proof is.

We do teach that Dafny assertions can be used as ``hints'' to the verifier checker; we also show how Dafny (ghost) functions can be used within specifications or assertions to embody lemmas that Dafny cannot find itself.  In the latter part of the course, questions require (ghost) data and methods to model the state of imperative objects. We mention Dafny's \texttt{calc} statement that supports line-by-line reasoning only in passing. 

We consider this a trade-off worth making: the course stays focused on program verification, through a programming lens, 
and we use the time to allow students to complete more significant examples with more complex verification constructs,
rather than teaching proof and necessarily working on smaller examples.

\section{Evaluation}

As part of VUW's quality assurance process, we conducted a standard
evaluation of SWEN324.  Under the terms of that process, we can
only report the quantitative results here.
The
quantitative questions employed a 5-point Likert scale (``Strongly
Agree/ Agree/ Neither Agree nor Disagree/ Strongly Disagree'' unless
otherwise noted) and employ both objective and affective questions. We
received 19 questionnaires from 88 students nominally enrolled in the
course when the evaluations where conducted.

Based on the quantitative feedback, over 70\% of students either agreed
or strongly agreed that the course was well organised, and that its
objectives were communicated well. 
70\% of students 
considered the workload ``about right'', although of the balance,
20\% considered the workload ``too much'' or ``far too much'' while only
5\% considered that that it was ``too little''.

Considering quality measures, most students considered the course
overall as ``very good'' (58\%) or ``excellent'' (21\%) --- although
one outlier did rank the course as ``poor''. Apart from that outlier,
all evaluated students agreed or strongly agreed that what they had
learned in the course had been valuable, and over half that the course
had stimulated interest in the subject ``a great deal''.  This results
in a median overall score or 2.0 ``very good''.  Compared with other
courses in the faculty, that is a slightly worse median (1.9), but
perhaps more relevant are comparisons with earlier offerings of more
traditional versions of the course. Over the last ten years, across
many iterations of the course, these have ranged from 3.8 ``Poor''
to 2.3 (approaching ``Good'') with most offerings around 2.6-2.7 ---
i.e.\ this version seems substantially better.

Finally, given the focus of our course design on online tools and
automatic marking to provide rapid feedback, it is gratifying that
80\% of students agreed or strongly agreed that the ``online
components of the course contributed to my learning''.  Over 90\%
agreed or strongly agreed that ``Assessment tasks have helped me to
learn'' and that ``I received helpful feedback on my progress.''  This
is about as strong evidence for the benefits of the ``programming
style'' approach we adopted in SWEN324, and the use of automated
marking and feedback, that one is ever likely to receive.

Overall we consider the experiment of our redesign of SWEN324 a success.  Following this programming-centric approach, almost all students were able to demonstrate enough engagement with practical software verification to pass the course, and those students who chose to put in the necessary time and effort were able to complete quite significant verification tasks.  In spite of the ``mastery'' approach taken in much of the course, the final assignments and essays, were sufficient to ensure 
a good spread of grades across the course.

We are aware that the practical, pragmatic, programming focus of this approach has some trade-offs and costs. While students are able to program with Dafny, their knowledge of logic and indeed of formal methods and software verification is latent, i.e.\ implicit.  For example, students would be able to propose preconditions for a given Dafny function (e.g.\ to avoid array bounds errors or invalid computation), and given interaction with a Dafny IDE, to write preconditions that Dafny could verify: many students could argue informally about why such preconditions were necessary.  Because the knowledge is not explicit, they would not be able to present the formal rationale for those preconditions, to derive them from e.g.\ weakest preconditions, or to produce a formal proof that those preconditions would definitively rule out crashes at run time.
We had hoped that these topics could be addressed in a 
follow-on fourth-year course, however it seems we will not have that opportunity.

The other costs were essentially resources: all students needed access to the Dafny tool at all times; 
technical support from tutors thus needed to be provided whenever possible.  Automated marking (both weekly questions and assignments) was essential to maintaining that programming focus, and directly supported learning. Preparing the automated questions, and then validating them by verifying several different solutions also required significant time
and effort, by both tutors and academic staff. Some of this
effort (e.g.\ weekly questions) could be amortised over
multiple offerings of the course, but most institutions
would need to refresh the main assignments for each course offering --- at least in institutions without very strong honour code traditions that prevent sharing solutions across cohorts. 
\section{Conclusion}
%\vspace*{-2mm}
\begin{verse}
  \textit{
Getting code to work is one thing.\\
Proving it does what it's supposed to is something else.\\
Convincing Dafny you've proved it does what it's supposed to\\
~~~~is something else entirely.}
\end{verse}
\vspace*{-10mm}
\begin{flushright}
``Motto for a Software Correctness Course''\\
Thomas J. ``Tad'' Peckish (attrib.), twitter, Oct 4 2020
\end{flushright}
%\vspace*{-1mm}

Formal methods are becoming more popular in software engineering
practice, and accordingly more common in software engineering
education course work. This shift has implications for how we teach: a
course that aims to ensure every computer science or software
engineering student has understanding of formal methods, and some
basic exposure to formal tools, must necessarily be different to a
course that (explicitly or implicitly) aims to prepare students for
graduate work.
We have
described our experience in redeveloping our formal methods course to
be for the many, not the few; by employing tool and strategies typically
used to teach programming, rather than those of mathematics.
So far, this approach has been fruitful: most students who enroll in
the course are able to pass it; are able to actually complete some
small problems using Dafny; and overall consider the course
worthwhile.
The key factors supporting this outcome were the Dafny tool, which is
now sufficiently mature to be used at this scale, and the necessary
time and effort to prepare weekly questions and assignments in advance
to support feedback via automatic marking.  We hope to continue with
work, both to integrate formal methods ever more tightly into teaching
programming, and to investigate how tools such as Dafny can best
support this approach.

%% sstudents get feedback
%% expectations are clear
%% can work local or remotely (COVID)
%% sthudents always nkow where they are
%% students can choose how much time and effort to invest
%% mastery means can demonstrate a ``taste'' of the topic 

\section*{Acknowledgements}

Thanks to Rustan Leino and James Wilcox for all their help with Dafny;
to our colleagues Marco Servetto for the ``marcotron'' weekly question
system, to Royce Brown, Christo Muller, and the ECS technical staff
for their support with the course automation; to Lindsay Groves,
longtime custodian of Formal Methods at VUW through various
iterations (COMP202, SWEN202, SWEN224, SWEN324); to the reviewers for their helpful comments; and above all to the
students who choose to stay with SWEN324 in spite of everything.

This work was supported in part by the Royal Society of New Zealand Marsden Fund Grant VUW1815, and by a gift from Agoric. 

\bibliographystyle{splncs04}
\bibliography{dafny}

\begin{thebibliography}{10}
\providecommand{\url}[1]{\texttt{#1}}
\providecommand{\urlprefix}{URL }
\providecommand{\doi}[1]{https://doi.org/#1}

\bibitem{Aceto2021}
Aceto, L., Ing{\'o}lfsd{\'o}ttir, A.: Introducing formal methods to first-year
  students in three intensive weeks. In: Formal Methods Teaching Workshop. pp.
  1--17. Springer (2021)

\bibitem{TomAngeloWherever}
Angelo, T.: A teacher's dozen—fourteen general research-based principles for
  improving higher learning. {AAHE} Bulletin  (1993)

\bibitem{Arlt2014}
Arlt, S., Rubio-Gonz{\'a}lez, C., R{\"u}mmer, P., Sch{\"a}f, M., Shankar, N.:
  The gradual verifier. In: NASA Formal Methods Symposium. pp. 313--327.
  Springer (2014)

\bibitem{Astrauskas2019}
Astrauskas, V., M{\"u}ller, P., Poli, F., Summers, A.J.: Leveraging rust types
  for modular specification and verification. Proceedings of the ACM on
  Programming Languages  \textbf{3}(OOPSLA),  1--30 (2019)

\bibitem{Bader2018}
Bader, J., Aldrich, J., Tanter, {\'E}.: Gradual program verification. In:
  VMCAI. pp. 25--46 (2018)

\bibitem{Beckert2012}
Beckert, B., Grebing, S.: Evaluating the usability of interactive verification
  systems. In: COMPARE. pp. 3--17. Citeseer (2012)

\bibitem{why3teaching2019}
Blazy, S.: Teaching deductive verification in {Why3} to undergraduate students.
  In: Formal Methods Teaching (FMTea) (2019)

\bibitem{bornholt2021using}
Bornholt, J., Joshi, R., Astrauskas, V., Cully, B., Kragl, B., Markle, S.,
  Sauri, K., Schleit, D., Slatton, G., Tasiran, S., et~al.: Using lightweight
  formal methods to validate a key-value storage node in amazon s3. In:
  Proceedings of the ACM SIGOPS 28th Symposium on Operating Systems Principles.
  pp. 836--850 (2021)

\bibitem{Brooks1987}
Brooks, F., Kugler, H.: No silver bullet. April (1987)

\bibitem{FMFUN19}
Cerone, A., Roggenbach, M. (eds.): Formal Methods – Fun for Everybody
  {(FMFun)}. Springer (2019)

\bibitem{Chipala2013}
Chlipala, A.: Certified Programming with Dependent Types: A Pragmatic
  Introduction to the {Coq} Proof Assistant. {MIT} Press (2013)

\bibitem{Christakis2016}
Christakis, M., Leino, K.R.M., M{\"u}ller, P., W{\"u}stholz, V.: Integrated
  environment for diagnosing verification errors. In: International Conference
  on Tools and Algorithms for the Construction and Analysis of Systems. pp.
  424--441. Springer (2016)

\bibitem{Cook2018}
Cook, B.: Formal reasoning about the security of {A}mazon {W}eb {S}ervices. In:
  International Conference on Computer Aided Verification. pp. 38--47. Springer
  (2018)

\bibitem{Dietl2011}
Dietl, W., Dietzel, S., Ernst, M.D., Mu{\c{s}}lu, K., Schiller, T.W.: Building
  and using pluggable type-checkers. In: Proceedings of the 33rd International
  Conference on Software Engineering. pp. 681--690 (2011)

\bibitem{Dongol2019}
Dongol, B., Petre, L., Smith, G.: Formal Methods Teaching: Third International
  Workshop and Tutorial, FMTea 2019, Held as Part of the Third World Congress
  on Formal Methods, FM 2019, Porto, Portugal, October 7, 2019, Proceedings,
  vol. 11758. Springer Nature (2019)

\bibitem{teachDafny2021}
Ettinger, R.: Lessons of formal program design in {D}afny. In: Formal Methods
  Teaching (FMTea) (2021)

\bibitem{ferreiraformal}
Ferreira, J.F., Mendes, A., Menghi, C.: Formal Methods Teaching {(FMTea)}.
  Springer Nature (2021)

\bibitem{Figueroa2018}
Figueroa, I., Garc{\'\i}a, B., Leger, P.: Towards progressive program
  verification in dafny. In: Proceedings of the XXII Brazilian Symposium on
  Programming Languages. pp. 90--97 (2018)

\bibitem{matrixOracle}
Flannery-Dailey, F., Wagner, R.L.: Wake up! {G}nosticism and {B}uddhism in the
  {M}atrix. Journal of Religion \& Film  \textbf{5}(2) (2001)

\bibitem{Garavel2020}
Garavel, H., Ter~Beek, M.H., Van De~Pol, J.: The 2020 expert survey on formal
  methods. In: International Conference on Formal Methods for Industrial
  Critical Systems. pp. 3--69. Springer (2020)

\bibitem{Grebing2020}
Grebing, S., Ulbrich, M.: Usability recommendations for user guidance in
  deductive program verification. In: Deductive Software Verification: Future
  Perspectives, pp. 261--284. Springer (2020)

\bibitem{Green1996}
Green, T.R.G., Petre, M.: Usability analysis of visual programming
  environments: a ‘cognitive dimensions’ framework. Journal of Visual
  Languages \& Computing  \textbf{7}(2),  131--174 (1996)

\bibitem{Greengard2021}
Greengard, S.: The Internet of Things. MIT press (2021)

\bibitem{teachFRAMA2021}
G{\"u}demann, M.: Online teaching of verification of {C} programs in applied
  computer science. In: Formal Methods Teaching (FMTea) (2021)

\bibitem{irondafny}
Hawblitzel, C., Howell, J., Kapritsos, M., Lorch, J.R., Parno, B., Roberts,
  M.L., Setty, S.T.V., Zill, B.: {IronFleet}: proving safety and liveness of
  practical distributed systems. Commun. {ACM}  \textbf{60}(7),  83--92 (2017)

\bibitem{spinbook}
Holzmann, G.J.: The {SPIN} Model Checker: Primer and Reference Manual.
  Addison-Wesley (2003)

\bibitem{alloybook}
Jackson, D.: Software Abstractions: Logic, Language, and Analysis. {MIT Press}
  (April 2006)

\bibitem{Jones2021}
Jones, C.B., Misra, J.: Theories of Programming: The Life and Works of Tony
  Hoare. Morgan \& Claypool (2021)

\bibitem{Jung2017}
Jung, R., Jourdan, J.H., Krebbers, R., Dreyer, D.: {RustBelt:} securing the
  foundations of the rust programming language. Proceedings of the ACM on
  Programming Languages  \textbf{2}({POPL}),  1--34 (2017)

\bibitem{kamburjan2021increasing}
Kamburjan, E., Gr{\"a}tz, L.: Increasing engagement with interactive
  visualization: Formal methods as serious games. In: Formal Methods Teaching
  Workshop. pp. 43--59. Springer (2021)

\bibitem{Klabnik2019}
Klabnik, S., Nichols, C.: The {R}ust Programming Language (Covers {R}ust 2018).
  No Starch Press (2019)

\bibitem{reliance2021}
K{\"o}rner, P., Krings, S.: Increasing student self-reliance and engagement in
  model-checking courses. In: Formal Methods Teaching (FMTea) (2021)

\bibitem{tlabook}
Lamport, L.: Specifying Systems: The {TLA+} Language and Tools for Hardware and
  Software Engineers. Pearson (2002)

\bibitem{Dafny3.0.0}
Leino, K.R.M.: Dafny 3.0.0 release,
  \url{\texttt{https://\-github.com/\-dafny-lang/\-dafny/\-releases/\-tag/\-v3.0.0}}

\bibitem{leino2013}
Leino, K.R.M.: Developing verified programs with {D}afny. In: 2013 35th
  International Conference on Software Engineering (ICSE). pp. 1488--1490. IEEE
  (2013)

\bibitem{dafny2020}
Leino, K.R.M.: Program Proofs. Available from Lulu.com (2020)

\bibitem{autoactive2010}
Leino, K.R.M., Moskal, M.: Usable auto-active verification. In: Usable
  Verification Workshop {(UV10)} (2010)

\bibitem{rustan1998extended}
Leino, K.R.M., Nelson, G.: An extended static checker for {M}odula-3. In:
  International Conference on Compiler Construction. pp. 302--305. Springer
  (1998)

\bibitem{Markstrum2010}
Markstrum, S., Marino, D., Esquivel, M., Millstein, T., Andreae, C., Noble, J.:
  Javacop: Declarative pluggable types for java. ACM Transactions on
  Programming Languages and Systems (TOPLAS)  \textbf{32}(2),  1--37 (2010)

\bibitem{touchOfClass}
Meyer, B.: Touch of Class. Springer (2009)

\bibitem{Muller2011}
M{\"u}ller, P., Ruskiewicz, J.N.: Using debuggers to understand failed
  verification attempts. In: International Symposium on Formal Methods. pp.
  73--87. Springer (2011)

\bibitem{formed08}
Noble, J., Pearce, D.J., Groves, L.: Introducing {A}lloy in a software
  modelling course. In: 1st Workshop on Formal Methods in Computer Science
  Education ({FORMED}) (2008)

\bibitem{Pang2018}
Pang, A., Anslow, C., Noble, J.: What programming languages do developers use?
  a theory of static vs dynamic language choice. In: 2018 IEEE Symposium on
  Visual Languages and Human-Centric Computing (VL/HCC). pp. 239--247. IEEE
  (2018)

\bibitem{CoqTute11}
Paulin{-}Mohring, C.: Introduction to the {Coq} proof-assistant for practical
  software verification. In: {LASER} International Summer School. Springer
  (2011)

\bibitem{Pearce2015}
Pearce, D.J., Groves, L.: Designing a verifying compiler: Lessons learned from
  developing {W}hiley. Science of Computer Programming  \textbf{113},  191--220
  (2015)

\bibitem{bladeRunnerFilm}
Scott, R.: Blade runner. Motion Picture (1982)

\bibitem{wayne2018temporal}
Wayne, H.: Temporal logic. In: Practical TLA+, pp. 97--110. Springer (2018)

\bibitem{Wise2020}
Wise, J., Bader, J., Wong, C., Aldrich, J., Tanter, {\'E}., Sunshine, J.:
  Gradual verification of recursive heap data structures. Proceedings of the
  ACM on Programming Languages  \textbf{4}(OOPSLA),  1--28 (2020)

\bibitem{Yatapanage2021}
Yatapanage, N.: Introducing formal methods to students who hate maths and
  struggle with programming. In: Formal Methods Teaching Workshop. pp.
  133--145. Springer (2021)

\bibitem{ZZZ}
Zhumagambetov, R.: Teaching formal methods in academia: A systematic literature
  review. In: Formal Methods – Fun for Everybody {(FMFun)} (2019)

\end{thebibliography}

\end{document}